\documentclass[sigconf]{acmart}
\usepackage{multirow}

\makeatletter
\newcommand{\thickhline}{%
    \noalign {\ifnum 0=`}\fi \hrule height 1pt
    \futurelet \reserved@a \@xhline
}
\makeatother

\def\BibTeX{{\rm B\kern-.05em{\sc i\kern-.025em b}\kern-.08emT\kern-.1667em\lower.7ex\hbox{E}\kern-.125emX}}
    
\copyrightyear{2019}
\acmYear{2019} 
\setcopyright{iw3c2w3}
\acmConference[WWW '19]{Proceedings of the 2019 World Wide Web Conference}{May 13--17, 2019}{San Francisco, CA, USA}
\acmBooktitle{Proceedings of the 2019 World Wide Web Conference (WWW '19), May 13--17, 2019, San Francisco, CA, USA}
\acmPrice{}
\acmDOI{10.1145/3308558.3313466}
\acmISBN{978-1-4503-6674-8/19/05}

\begin{document}

%
% The "title" command has an optional parameter, allowing the author to define a "short title" to be used in page headers.
\title{Learning Fast Matching Models from Weak Annotations}

%
% The "author" command and its associated commands are used to define the authors and their affiliations.
% Of note is the shared affiliation of the first two authors, and the "authornote" and "authornotemark" commands
% used to denote shared contribution to the research.
\author{Xue Li}
\authornote{Both authors contributed equally to this work.}
\affiliation{Microsoft}
\email{xeli@microsoft.com}

\author{Zhipeng Luo}
\authornotemark[1]
\affiliation{Microsoft}
\email{Zhipeng.Luo@microsoft.com}

\author{Hao Sun}
\affiliation{Microsoft}
\email{hasun@microsoft.com}

\author{Jianjin Zhang}
\authornote{This work is conducted during internship at Microsoft.}
\affiliation{Microsoft}
\affiliation{Tsinghua University}
\email{swjxzjj@gmail.com}

\author{Weihao Han}
\affiliation{Microsoft}
\email{weihan@microsoft.com}

\author{Xianqi Chu}
\affiliation{Microsoft}
\email{Xianqi.Chu@microsoft.com}

\author{Liangjie Zhang}
\affiliation{Microsoft}
\email{liazha@microsoft.com}

\author{Qi Zhang}
\affiliation{Microsoft}
\email{Zhang.Qi@microsoft.com}

\renewcommand{\shortauthors}{Xue Li, et al.}

%
% By default, the full list of authors will be used in the page headers. Often, this list is too long, and will overlap
% other information printed in the page headers. This command allows the author to define a more concise list
% of authors' names for this purpose.

%
% The abstract is a short summary of the work to be presented in the article.
\begin{abstract}
We propose a novel training scheme for fast matching models in Search Ads, motivated by practical challenges. The first challenge stems from the pursuit of high throughput, which prohibits the deployment of inseparable architectures, and hence greatly limits model accuracy. The second problem arises from the heavy dependency on human provided labels, which are expensive and time-consuming to collect, yet how to leverage unlabeled search log data is rarely studied. The proposed training framework targets on mitigating both issues, by treating the stronger but undeployable models as annotators, and learning a deployable model from both human provided relevance labels and weakly annotated search log data. Specifically, we first construct multiple auxiliary tasks from the enumerated relevance labels, and train the annotators by jointly learning from those related tasks. The annotation models are then used to assign scores to both labeled and unlabeled training samples. The deployable model is firstly learnt on the scored unlabeled data, and then fine-tuned on scored labeled data, by leveraging both labels and scores via minimizing the proposed label-aware weighted loss. According to our experiments, compared with the baseline that directly learns from relevance labels, training by the proposed framework outperforms it by a large margin, and improves data efficiency substantially by dispensing with 80\% labeled samples. The proposed framework allows us to improve the fast matching model by learning from stronger annotators while keeping its architecture unchanged. Meanwhile, it offers a principled manner to leverage search log data in the training phase, which could effectively alleviate our dependency on human provided labels.
\end{abstract}



%
% Keywords. The author(s) should pick words that accurately describe the work being
% presented. Separate the keywords with commas.

%
% A "teaser" image appears between the author and affiliation information and the body 
% of the document, and typically spans the page. 
%   \Description{Enjoying the baseball game from the third-base seats. Ichiro Suzuki preparing to bat.}

%
% This command processes the author and affiliation and title information and builds
% the first part of the formatted document.
\maketitle

% ---------------------------------------------------------------------------------

\section{Introduction}
\label{sec.introduction}

The recent decade has witnessed a booming of sponsored search due to the outburst of search demands, which raises severe challenges to the design of practical search algorithms.
As a starting step of typical sponsored search systems, the fast matching model discussed here is designed to recall from the entire ad corpus a list of candidates best matching a given query. All the subsequent processing steps will be applied on the selected list only. The extremely intensive user requests of Search Ads system poses stringent online latency requirement on the fast matching model.
To some extent, its accuracy and latency forms the upper bound of the overall performance. To improve the fast matching model, we need either to leverage more powerful model architectures, or more effective training strategies.
However, both directions turn out to be difficult, and will suffer from restrictions from both system and data.

The first restriction relates to model architecture.
Commercial search engines need to handle extremely intensive user requests by searching from a daunting ad corpus, and their pursuit of high throughput prohibits the deployment of inseparable model architectures, especially those involving heavy online computations.
For this reason, relevance model is usually implemented as Siamese networks such as CDSSM~\cite{shen2014learning}, which transforms the input query and ad individually to representing vectors residing in a common inner-product space, without knowledge of each other.
Such an architecture allows us to reduce online computations following a space-for-time strategy, by hosting online a dictionary of pre-computed ad vectors.
When a query is issued, the model simply retrieves relevant ad vectors by matching it from the dictionary, and the matching could be effectively done by resorting to approximate nearest neighbor search algorithms such as NGS~\cite{wang2012query}.
However, due to the deferment of interactions between queries and ads, Siamese networks often suffer from losing details important to the matching task in the forward pass~\cite{hu2014convolutional}, which greatly limits matching performance. Note that better accuracy could also be achieved using more complex model architectures, especially those involving interactions between query and ad such as Deep Crossing, or model ensembles involving crossing features. However, such models are mostly inseparable, hence could not be deployed in the same way as Siamese networks.

The second restriction comes from the training data.
Relevance model is typically trained on human provided labels, which is expensive and time-consuming to collect.
In contrast to this, the search log contains tremendous amount of data, but relevance labels are unavailable.
An alternative solution is to employ user clicks as a surrogate of relevance labels, by taking clicked $<$\textit{query}, \textit{ad}$>$ pair as positive, and synthesizing negative pairs by fixing a query while randomly sampling ads from the same training batch.
This strategy oversimplifies the learning task by transforming matching to a classification problem~\cite{wu018learning}, and introduces ambiguities from three aspects.
First of all, the arbitrariness and subjectivity of user behavior leads to a misalignment between user clicks and true relevance labels, polluting the training set with false positives.
Secondly, data synthesizing may introduce false negatives, and most of the synthetic negative pairs share no common terms in queries and ads.
Learning on such data may mislead the model to consider common terms as critical evidences of relevance, whereas lexically similar query and ad may have totally different intents, such as \textit{iphone} and \textit{iphone cover}.
Last but not least, human labels typically have enumerated values to distinguish relevance with finer granularity, which are difficult to approximate by binary-valued clicks.

This paper targets on mitigating both the aforementioned issues.
For this purpose, we employ powerful but undeployable models to assign weak annotations or scores to unlabeled data, and improve the deployable model by learning from both the weak annotations and true human labels.
A schematic illustration of the proposed framework is plotted in Figure~\ref{fig.overview}, which consists of four major steps.
Specifically, we first construct auxiliary tasks using enumerated relevance labels, and train annotators by jointly learning multiple tasks, following the idea of Multi-task Learning~(MTL).
The learnt annotators are then used to assign scores to both unlabeled and labeled samples.
Based on this, a deployable model is trained firstly on the scored unlabeled data, and then fine-tuned on scored labeled data.
A label-aware weighted loss function is proposed in the last step to leverage both labels and scores.

The advantages of the proposed framework are twofold.
At the foremost, it offers us great flexibility in the architecture design of fast matching model, since we can improve it by learning from stronger annotators, while keeping the model to be deployed as efficient as possible.
Because the annotation model runs offline only, it is allowed to explore complicated model architectures involving heavy computations, and is able to leverage either policy-driven rules or features that might be too heavy to access for online processing.
Secondly, this framework provides a principled way to leverage unlabeled data for learning fast matching models, which dispenses entirely with user clicks to avoid data pollution.
%Secondly, the proposed framework dispenses with user clicks, and is designed to leverage data that are totally unlabeled.
%This not only avoids data pollution, but also allows us to collect as many training data as we want.

According to our experiments, compared with training directly on relevance labels, the proposed framework is able to improve \textit{ROC AUC} by 4.5\% and \textit{PR AUC} by 3.2\%, demonstrating the effectiveness of improving relevance model by leveraging unlabeled data and stronger annotators.
In addition, with only 20\% labeled samples, we are able to achieve performance comparable to the baseline trained on the entire labeled set, which effectively alleviates our dependency on human provided labels.

Our main contributions are summarized as follows.
\begin{itemize}
    \item we propose a novel training framework for the learning of fast matching models in search ads, which is able to leverage more powerful but undeployable models as weak annotators, and learn a simple and deployable model from both the weak annotations and the true human labels. The weak annotators are trained following MTL scheme to fully leverage the fine-grained relevance labels, and a label-aware weighted loss is proposed to softly combine the weak annotations and human provided labels.
    \item we evaluate the proposed training framework on large datasets, where both \textit{PR AUC} and \textit{ROC AUC} are significantly improved, demonstrating its effectiveness. Our experiments also shows that training by the proposed framework could greatly reduce our dependency on human labels.
\end{itemize}

% --------------------------------------------------------------------------------

\section{Related Work}
\label{sec.relatedwork}

The proposed framework is mainly fueled by the recent progress in web search and model compression.

Existing methods on semantically matching queries and documents could be roughly grouped into two categories, namely traditional approaches and deep learning based models. Representative methods falling into the former category include LSA~\cite{deerwester1990lsa}, pLSA~\cite{hofmann1999plsa}, topic models such as LDA~\cite{blei2003latent}, Bi-Lingual Topic Models~\cite{gao2011clickthrough}, etc.
Methods belonging to the latter category are designed to extract semantic information via deep learning architectures, including auto-encoders~\cite{salakhutdinov2009semantic}, Siamese networks~\cite{huang2013learning,shen2014learning,shen2014latent,gao2014modeling,hu2014convolutional,tai2015improved,severyn2015learning}, interaction-based networks~\cite{hu2014convolutional,lu2013advances,wan2016deep,yang2016anmm,yin2016abcnn}, and lexical and semantic matching networks~\cite{mitra2017learning,guo2016deep}, etc. Readers may refer to~\cite{mitra2017neural} for a tutorial on this topic.
The framework proposed in this paper adopts CDSSM~\cite{shen2014learning} and Deep Crossing~\cite{shan2016deep} as key building blocks.
% , since the Siamese structure of CDSSM offers a convenient solution to reducing online computations.
However, instead of improving its architecture, we alternatively attempt to improve its training scheme in this paper.

The proposed framework also draws inspirations from the teacher-student scheme in model compression.
%Following another line of research, the teacher-student framework is initially developed as a branch of model compression, which targets on distilling knowledge from a large deep model into a light-weight one.
As an initial attempt, the authors of~\cite{bucilua2006model} train a smaller model using synthetic data labeled by a teacher model. In~\cite{ba2014do}, the authors propose to train the student model by mimicking the logit values of the teacher model, instead of directly learning the softmax outputs. This is extended later in~\cite{romero2014}, by incorporating intermediate hidden layer outputs as target values as well. Hinton et al.~\cite{hinton2015distilling} propose to adjust the temperature variable in the softmax function. Recently the authors of~\cite{bharat2016deep} introduce a noise-based regularizer while training student models. In this paper, the teacher-student framework is adapted in several aspects to improve relevance algorithms in sponsored search.

While writing this paper, we also notice a recent work in~\cite{wu018learning}, which extends the teacher-student scheme to the learning of retrieval-based chatbots, where a sequence-to-sequence model is employed as a weak annotator, and the student model is trained by learning from both the weak signals and the unlabeled data.
%The proposed framework is instantiated using CDSSM~\cite{shen2014learning} as student model and Deep Crossing~\cite{shan2016deep} as teacher model.
Both \cite{wu018learning} and our proposed framework are motivated by the success of teacher-student scheme, but we adapt it in different directions to align with our respective task.
% As shown in Section~\ref{sec.exp}, the shift of application scenario not only inspires task-driven adaptations, but also derives some illuminating experimental results.

% ----------------------------------------------------------------------------------

\section{Proposed Framework}
\label{sec.method}

\begin{figure*}[!t]
\begin{center}
   \includegraphics[width=0.6\textwidth]{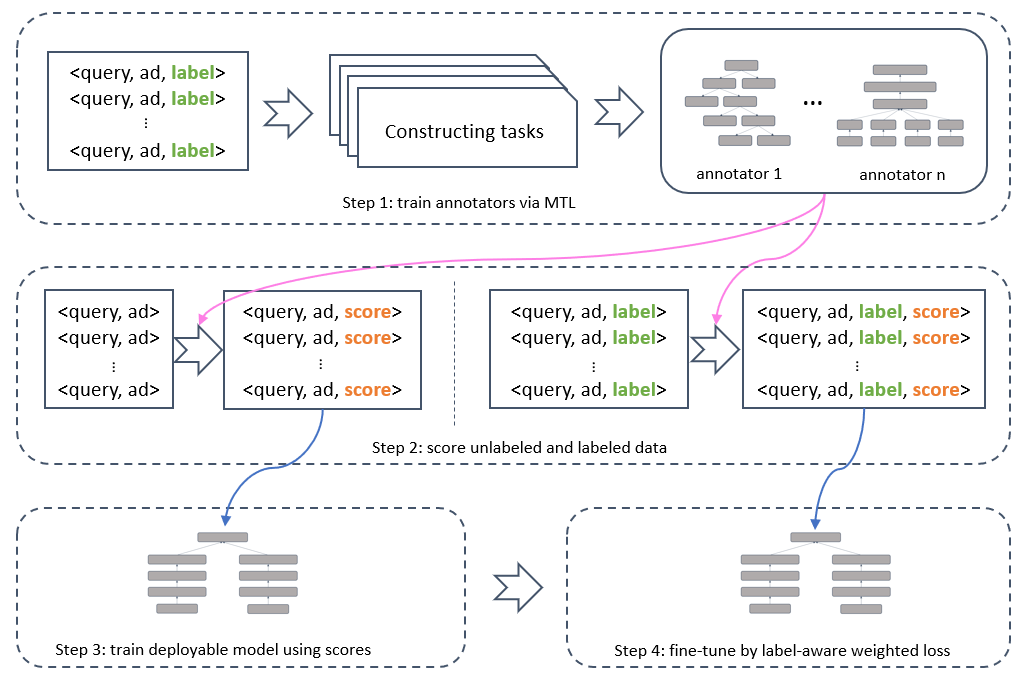}
\end{center}
   \caption{Illustration of the proposed framework, which includes four steps. The first step constructs multiple related tasks, and trains annotation models using labeled data, via MTL. The second step scores both the labeled and unlabeled data using the annotators. A deployable model is trained on the scored unlabeled data in the third step. In the last step, the deployable model is fine-tuned by minimizing the label-aware weighted loss. All steps involve offline computations only.}
   \label{fig.overview}
\end{figure*}

% This section begins with a brief introduction of sponsored search basics, followed by a detailed explanation of the proposed framework.

\subsection{Overview}
\label{sec.method.overview}

\subsubsection{Sponsored Search}
\label{sec.method.overview.term}

Below we list some useful terminologies, and we refer readers to~\cite{shan2016deep} for more details.

\textit{Impression} is an instance of an ad displayed to a user.
\textit{Click} indicates whether an impression is clicked by a user.
\textit{Click-Through Rate} or \textit{CTR} is the ratio of the total number of clicks to total number of impressions.
\textit{Query} is a text string issued by a user in a search engine.
\textit{Keyword} is a text string concerning a certain product, specified by an advertiser.
\textit{Landing Page} or \textit{LP} refers to the web page a user reaches after clicking the corresponding ad of a product.
A brief description of the product is usually displayed as the title of the \textit{LP}, called \textit{LP} title.
\textit{Ad Copy} or \textit{AC} refers to the content seen by a user before clicking, which typically includes ad title, description, displayed URL, landing page link, etc.
\textit{Listing} contains the bided keywords associated with Ad Copy, bid, etc.
\textit{Relevance labels} are a set of predefined judgements used to characterize the relevance of query and ad listing.
We have two label sets, namely \textit{AC} labels that measure the relevance between query and Ad Copy contents, and \textit{LP} labels that measure relevance between query and landing page contents.
Throughout this paper, our label set is \{0, 1, 2, 3, 4\} for \textit{AC} labels, and \{0, 1, 2, 3, 4, 5\} for \textit{LP} labels, with higher labels indicating better relevance.

Our system is akin to the one described in~\cite{ling2017model}, which consists of several steps including \textit{selection}, \textit{relevance filtration}, \textit{CTR prediction}, \textit{ranking} and \textit{allocation}.
% In particular, the \textit{selection} step is designed to fast recall from the entire listing corpus a list of candidates most relevant to the given query.
% Our discussions in this paper focuses on the relevance task in \textit{selection}, i.e. evaluating relevance between query and ad listing in the \textit{selection} step.
% Unlike \textit{relevance filtration} that operates on the selected candidates only, our task needs to consider a vast quantity of listings, making latency a major concern.

\subsubsection{Proposed Framework}
\label{sec.method.overview.outline}

Figure~\ref{fig.overview} sketches the main idea of the proposed framework, which contains four major steps.
The first step trains annotation models on human labeled data, by jointly learning multiple related tasks.
The second step scores both the labeled and unlabeled data using this annotation model.
In the following step, a deployable model is trained on the scored unlabeled data.
Finally, the deployable model is fine-tuned on both the labels and scores of the labeled data, by minimizing the label-aware weighted loss.
All steps involve purely offline computations; only the final simple model will be deployed online.
%The details of several major steps will be explained in the following subsections.
We instantiate our deployable model as CDSSM~\cite{shen2014latent}, and implement two annotation models, i.e. Deep Crossing~(abbreviated as DC) in~\cite{shan2016deep} and a decision tree ensemble model~(abbreviated as DT) in~\cite{ling2017model}.
% CDSSM is a typical Siamese network that well suits our task, while both Deep Crossing and decision tree model have very simple structures.
% We would like to show that learning by the proposed framework could improve CDSSM considerably, even with very simple annotators.

\subsection{Annotation Model Training via MTL}
\label{sec.method.teacher}

\begin{figure}[!t]
\begin{center}
   \includegraphics[width=0.5\textwidth]{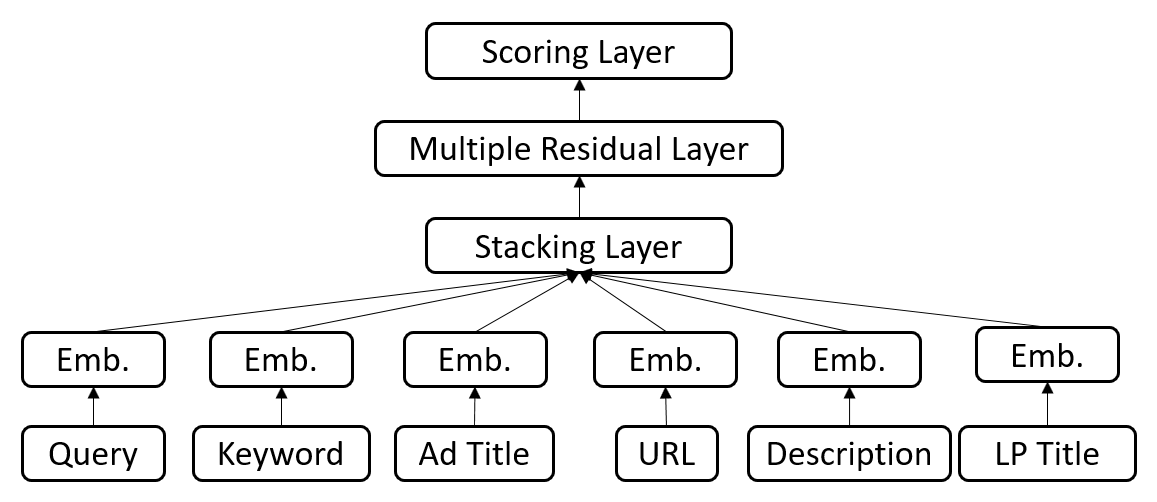}
\end{center}
   \caption{Architecture of Deep Crossing, which is inseparable and cannot be deployed as a fast matching model.}
   \label{fig.dc}
\end{figure}

Two annotation models are considered in this paper, i.e. Deep Crossing and decision tree ensemble.
Figure~\ref{fig.dc} shows the structure of Deep Crossing, which contains an input layer, an embedding layer, multiple residual layers, and a scoring layer. Deep Crossing introduces feature interactions immediately after embedding layers, hence is inseparable.
Our decision tree ensemble model follows the one described in~\cite{ling2017model}, which takes as input a set of features, some of which are difficult to access in the recall phase.
Therefore, both annotation models could not be deployed as fast matching models directly.
Annotation models are trained on labeled data, where each sample is a triple of query, ad listing and human judged relevance label chosen from a predefined label set, see Section~\ref{sec.method.overview}.
%Our label set is \{\textit{perfect}, \textit{excellent}, \textit{good}, \textit{fair}, \textit{bad}\} for AC labels, and \{\textit{perfect}, \textit{excellent}, \textit{good}, \textit{fair}, \textit{bad}, \textit{very bad}\} for LP labels.
% The most trivial way of feeding the labeled samples to annotation models is to consider samples with \textit{AC} or \textit{LP} label 0 as negative, while others as positive.
% However, this will obviously weaken the discriminative power of labels.

We construct several auxiliary classification tasks by partitioning the label sets with different pivots, as shown in Table~\ref{table.auxiliary}, where four auxiliary tasks are constructed for \textit{LP} labels, and three for \textit{AC} labels, along with the main task. The final loss is calculated as a weighted sum of the individual loss for each task, with the main task weighted as $0.5$ while all the auxiliary tasks evenly weighted. The final score is computed analogously.

% Jointly learning multiple related tasks enables our annotation model to distinguish relevance with finer granularity using shared parameters, which is critical for relevance models since we need to rank listings by their relevance to a given query.

\begin{table}[h]
\caption{Constructing auxiliary tasks.}
    \begin{center}
    \begin{tabular}{c|l|l}
        \thickhline
        Label Set                       &Task                           &Negative Labels\\
        \hline %\hline
        \textit{AC}                     &\multirow{2}{*}{main task}     &\{0\}\\
        \cline{1-1} \cline{3-3}
        \textit{LP}                     &                               &\{0\}\\
        \hline %\hline
        \multirow{3}{*}{\textit{AC}}    &aux. task 1                    &\{0, 1\}\\
        %\cline{2-3}
                                        &aux. task 2                    &\{0, 1, 2\}\\
        %\cline{2-3}
                                        &aux. task 3                    &\{0, 1, 2, 3\}\\
        \hline
        \multirow{4}{*}{\textit{LP}}    &aux. task 1                    &\{0, 1\}\\
        %\cline{2-3}
                                        &aux. task 2                    &\{0, 1, 2\}\\
        %\cline{2-3}
                                        &aux. task 3                    &\{0, 1, 2, 3\}\\
        %\cline{2-3}
                                        &aux. task 4                    &\{0, 1, 2, 3, 4\}\\
        \thickhline
    \end{tabular}
    \end{center}

    \label{table.auxiliary}
\end{table}

\subsection{CDSSM Training using Scores}
\label{sec.method.student}

\begin{figure}[!t]
\begin{center}
   \includegraphics[width=0.5\textwidth]{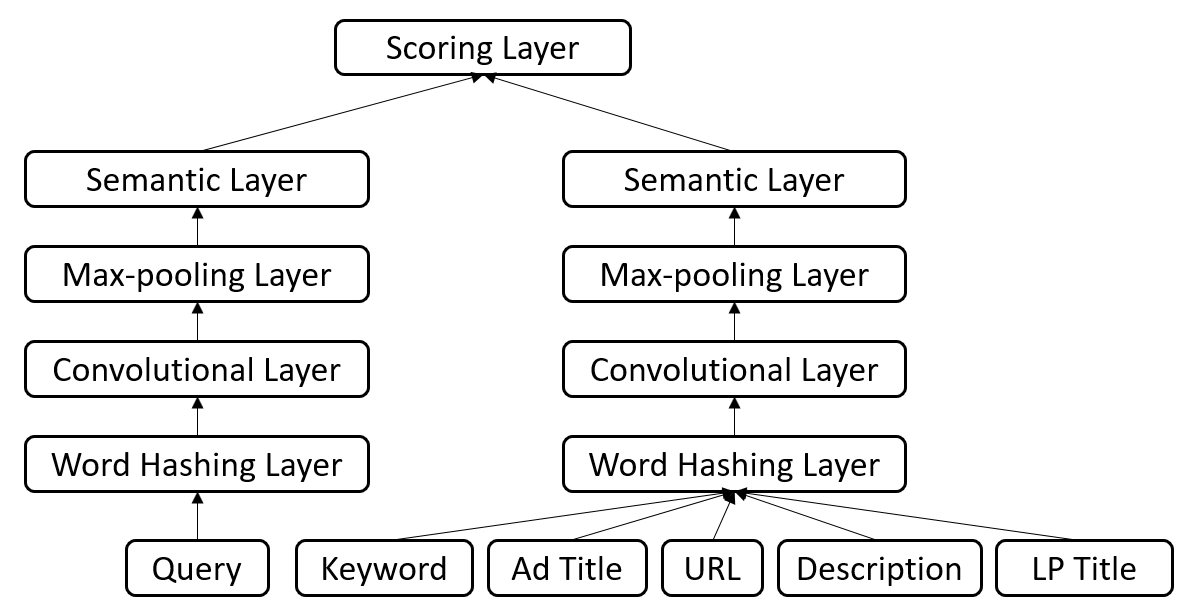}
\end{center}
   \caption{Architecture of CDSSM, which is separable and takes identical inputs with Deep Crossing.}
   \label{fig.cdssm}
\end{figure}

The architecture of CDSSM is shown in Figure~\ref{fig.cdssm}, which contains an input layer, a word hashing layer to transform each word into a count vector of letter-tri-grams, a convolutional layer to extract sliding window based contextual features around each word, a max-pooling layer to summarize most salient contextual features, followed by a semantic layer and the final cosine scoring layer.
%To facilitate experimentation, we use identical input features for Deep Crossing and CDSSM.

The scores from the annotation models could be leveraged either in a hard manner by thresholding them to binary values, or in a soft way by maintaining their original values. In addition, the reliability of the scores also needs investigation.
For this purpose, we define the targets and weights of the unlabeled samples as functions of annotation model scores:
\begin{eqnarray}\label{eqn.label}
y_i & = & f(s_i),\label{eqn.f}\\
\omega_i & = & g(s_i),\label{eqn.g}
\end{eqnarray}
where $y_i$, $\omega_i$ and $s_i$ are the target, weight, and annotation model score for the $i$th sample in the unlabeled dataset, respectively, while $f(\cdot)$ and $g(\cdot)$ are mapping functions from scores to targets and weights. When multiple annotation models are available, $s_i$ is calculated as their arithmetic mean.
Following this definition, we define several mapping functions as below:

\begin{eqnarray}
    f_1(s_i) & = & \left\{
               \begin{array}{ll}
                 1, & 0.5 \leq s_i < 1;\\
                 0, & 0 < s_i < 0.5;
               \end{array}
             \right.\label{eqn.f1}\\
    f_2(s_i) & = & s_i,\label{eqn.f2}\\
    g_1(s_i) & = & \left\{
               \begin{array}{ll}
                 0, & t_1 < s_i < t_2;\\
                 1, & \textrm{o.w.}
               \end{array}
             \right.\label{eqn.g1}\\
    g_2(s_i) & = & \big|2s_i - 1\big|^p,\label{eqn.g2}\\
    g_3(s_i) & = & 1.\label{eqn.g3}
\end{eqnarray}

Equation~(\ref{eqn.f1}) defines a hard target mapping, and Equation~(\ref{eqn.f2}) defines a soft one.
Equation~(\ref{eqn.g1}) to Equation~(\ref{eqn.g3}) define three weight mappings.
In implementation, Equation~(\ref{eqn.f1}) is trained by minimizing cross-entropy loss, while Equation~(\ref{eqn.f2}) by weighted mean square error~(MSE) loss.
Specifically, denoting the predicted score from CDSSM as $\hat{y}_i$ for the $i$th sample, we define its sample loss as below:
\begin{eqnarray}
\ell_i  =  \omega_i \cdot (y_i - \hat{y}_i)^2.\label{eqn.loss1}
\end{eqnarray}

\subsection{Fine-tuning by Label-aware Weighted Loss}
\label{sec.method.finetune}

% The labeled samples now are associated with both , both should be leveraged in the final step.

\begin{figure}[!t]
\begin{center}
   \includegraphics[width=0.5\textwidth]{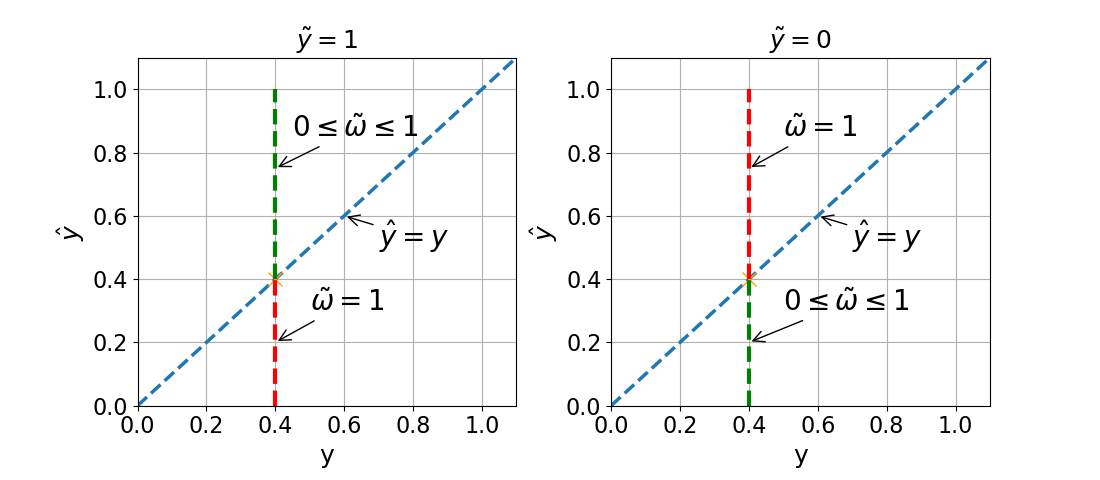}
\end{center}
   \caption{Illustration of the label-aware weights, where $\tilde{y}$ is the binary label, $y$ is the target from annotation model scores, and $\hat{y}$ is the prediction of CDSSM. When $\tilde{y}_i = 1$~(left), $\hat{y}_i > y_i$ appears more acceptable than $\hat{y}_i < y_i$, and hence is assigned a smaller weight, as highlighted by green. The right part shows $\tilde{y}_i = 0$, where $\hat{y}_i < y_i$ is given a smaller weight.}
   \label{fig.loss}
\end{figure}

To balance human judged relevance labels and annotation model scores, we propose the label-aware weights, as illustrated in Figure~\ref{fig.loss}. Denote the binary relevance labels as $\tilde{y}_i$ for sample $i$, where the binary labels are defined according to the main task in Table~\ref{table.auxiliary}. As shown in the left of Figure~\ref{fig.loss}, $\tilde{y}_i = 1$ indicates that sample $i$ has positive relevance labels, making it more acceptable when the predicted score $\hat{y}_i$ is greater than the target value $y_i$ obtained from annotation models, and less acceptable when $\hat{y}_i < y_i$. To reflect such differences, we define a weight $\tilde{\omega}_i$ for sample $i$, with $\tilde{\omega}_i < 1$ for acceptable loss and $\tilde{\omega}_i = 1$ otherwise. The right part of Figure~\ref{fig.loss} shows the opposite case, where $\tilde{y}_i = 0$ makes $\hat{y}_i > y_i$ less acceptable.

To facilitate the formulation of label-aware weights, we first define a sign function as shown below:
\begin{eqnarray}\label{eqn.omega}
\delta_\theta(x)  =   \left\{
                \begin{array}{ll}
                 \theta, & \textrm{if} \hspace{1mm} x \leq 0 ;\\
                 1, & \textrm{if} \hspace{1mm} x > 0,
               \end{array}
            \right.
\end{eqnarray}
where $0 \leq \theta \leq 1$ is a hyper-parameter indicating the discount of punishment imposed on acceptable loss. Based on the sign function in Equation~(\ref{eqn.omega}), our label-aware weighted loss could be defined as follows:
\begin{eqnarray}\label{eqn.loss2}
\tilde{\ell}_i
&=& \delta_\theta(y_i-\hat{y}_i)^{\tilde{y}_i} \cdot \big(\theta + 1 - \delta_\theta(y_i-\hat{y}_i)\big)^{(1 - \tilde{y}_i)} \cdot   \nonumber \\
&& (y_i-\hat{y}_i)^2 \nonumber \\
& \triangleq & \tilde{\omega}_i \cdot (y_i-\hat{y}_i)^2,
\end{eqnarray}
where label-aware weight $\tilde{\omega}_i$ is a function of $y_i$, $\tilde{y}_i$ and $\hat{y}_i$.

% Compared with Equation~(\ref{eqn.loss1}), Equation~(\ref{eqn.loss2}) emphasizes more on samples with less acceptable loss, with the degree of emphasis controlled by $\theta$. 
% More precisely, no punishments are imposed on acceptable loss when $\theta = 0$, and Equation~(\ref{eqn.loss2}) reduces to Equation~(\ref{eqn.loss1}) when $\theta = 1$.

% -----------------------------------------------------------------------------------------------------------------------------------------------------------------------------------------

\section{Experiments}
\label{sec.exp}

% This section attempts to examine that 1) whether training by the proposed framework is beneficial, and 2) the influence of annotation models, mapping functions as well as label-aware weights.

\subsection{Experimental Setup}
\label{sec.exp.setup}

\subsubsection{Datasets}
\label{sec.exp.setup.datasets}

Validating the proposed framework needs three datasets: a labeled one with human judged relevance labels~(referred to as labeled), an unlabeled one with both clicked and unclicked samples~(referred to as unlabeled), and an unlabeled one with only clicked samples~(referred to as clicked). The unlabeled and clicked datasets are sampled from the search log of a commercial search engine. A subset with comparable size to the test set is randomly sampled from the training labeled set for validation.
Statistics of the three datasets are shown in Table~\ref{table.dataset}.

\begin{table}[!t]
\caption{Statistics of datasets.}
    \begin{center}
    \begin{tabular}{l|l|l|l|l}
        \thickhline
        \multirow{2}{*}{}   &\multirow{2}{*}{Unlabeled}     &\multirow{2}{*}{Clicked}       &\multicolumn{2}{c}{Labeled}\\
        \cline{4-5}
                            &                               &                               &positive           &negative\\
        \hline
        train               &159,352 K                      &168,671 K                      &5,115 K            &9,752 K\\
        test                &--                             &--                             &132 K              &120 K\\
        \thickhline
    \end{tabular}
    \end{center}

    \label{table.dataset}
\end{table}

\subsubsection{Implementation}
\label{sec.exp.setup.implement}

We implement Deep Crossing following the description of~\cite{shan2016deep}, with several modifications including adding Batch Normalization~\cite{sergey2015batch} and soft-aligning query and ad using the neural attention similar to~\cite{bahdanau2014neural,parikh2016decomposable}, by concatenating ad inputs into a single string. 
% The dimension of the embedding layer is set to 128, and 3 residual layers are used, each has a dimension of 256.
We use identical CDSSM architecture in the last two steps in Figure~\ref{fig.overview}. 
% As for hyper-parameters, we set $\theta = 0.5$ when fine-tuning by the proposed label-aware weighted loss, and $t_1 = 0.25, t_2 = 0.75$ when implementing Equation~(\ref{eqn.g1}).
we conduct evaluations using \textit{ROC AUC} and \textit{PR AUC} as measurements, which represent the area under the Receiver Operating Characteristic curve and the Precision-Recall curve, respectively.
% Specifically, \textit{ROC AUC} could be interpreted as the expected probability of ranking a randomly chosen positive sample in front of a randomly chosen negative one, while \textit{PR AUC} could better distinguish the precision for retrieval algorithms, especially when the number of negative samples dwarfs that of the positive ones. Both metrics are important when measuring the performance of relevance algorithms.

\subsection{Performance}
\label{sec.exp.perf}

\begin{table}[!t]
\caption{Comparison of performance.}
    \begin{center}
    \begin{tabular}{l|c|c|c}
        \thickhline
        Annotator        &CDSSM       &\textit{ROC AUC} &\textit{PR AUC}\\
        %\cline{3-4}
        %                                &                               &\textit{ROC}   &\textit{PR}\\
        \hline
        \multirow{2}{*}{--}             &CDSSM-click                    &0.7185         &0.7890\\
                                        &CDSSM-labeled                  &0.8131         &0.8711\\
        \hline
        DC                              &\multirow{3}{*}{--}            &0.8425         &0.8946\\
        DT                              &                               &0.8448         &0.8940\\
        DC + DT                         &                               &0.8578         &0.9034\\
        \hline
        DC                              &\multirow{3}{*}{CDSSM}         &0.7937         &0.8625\\
        DT                              &                               &0.7936         &0.8604\\
        DC + DT                         &                               &0.8155         &0.8753\\
        \hline
        DC                              &\multirow{3}{*}{CDSSM-ft}      &0.8394         &0.8939\\
        DT                              &                               &0.8442         &0.8941\\
        DC + DT                         &                               &\bf{0.8497}    &\bf{0.8990}\\
        \thickhline
    \end{tabular}
    \end{center}

    \label{table.comparison}
\end{table}

\begin{table*}[!t]
\caption{Influence of annotation models and mapping functions.}
    \begin{center}
    \begin{tabular}{l|c|c|c|c|c|c|c|c|c}
    \thickhline
    \multirow{3}{*}{Annotator}    &\multicolumn{2}{c}{\multirow{2}{*}{\textit{AUC}-annotator}}  &\multicolumn{7}{|c}{\textit{AUC}-CDSSM}\\
    \cline{4-10}
                                &\multicolumn{2}{c|}{}   &before/after   &\multicolumn{2}{|c}{$f_1$ $g_1$}  &\multicolumn{2}{|c}{$f_1$ $g_2$}     &\multicolumn{2}{|c}{$f_2$ $g_3$}\\
    \cline{2-3} \cline{5-10}
                                &\textit{ROC}    &\textit{PR} &fine-tune   &\textit{ROC}    &\textit{PR} &\textit{ROC}    &\textit{PR} &\textit{ROC}    &\textit{PR}\\
    \hline
    \multirow{2}{*}{DC}         &\multirow{2}{*}{0.8046}    &\multirow{2}{*}{0.8659}    &before      &0.7164     &0.8146     &0.7273     &0.8186     &0.7143     &0.8116\\
                                &                           &                           &after       &0.7616     &0.8341     &0.7591     &0.8328     &0.7567     &0.8304\\
    \multirow{2}{*}{DC-BN}      &\multirow{2}{*}{0.8328}    &\multirow{2}{*}{0.8877}    &before      &0.7704     &0.8464     &0.7759     &0.8496     &0.7778     &0.8517\\
                                &                           &                           &after       &0.8219     &0.8773     &0.8262     &0.8814     &0.8301     &0.8841\\
    \multirow{2}{*}{DC-BN-A}    &\multirow{2}{*}{0.8351}    &\multirow{2}{*}{0.8886}    &before      &0.7744     &0.8502     &0.7797     &0.8524     &0.7862     &0.8576\\
                                &                           &                           &after       &0.8254     &0.8838     &0.8268     &0.8848     &0.8330     &0.8887\\
    \multirow{2}{*}{DC-BN-MTL}  &\multirow{2}{*}{0.8391}    &\multirow{2}{*}{0.8932}    &before      &0.7823     &0.8565     &0.7847     &0.8578     &0.7890     &0.8605\\
                                &                           &                           &after       &0.8277     &0.8857     &0.8296     &0.8868     &0.8343     &0.8896\\
    \multirow{2}{*}{DC-BN-A-MTL}&\multirow{2}{*}{0.8425}    &\multirow{2}{*}{0.8946}    &before      &0.7761     &0.8527     &0.7851     &0.8571     &0.7937     &0.8625\\
                                &                           &                           &after       &0.8281     &0.8863     &0.8309     &0.8887     &0.8372     &0.8922\\
    \thickhline
    \end{tabular}
    \end{center}

    \label{table.analyze}
\end{table*}

Table~\ref{table.comparison} summarizes our main results, where DC and DT represent Deep Crossing and decision tree ensemble, respectively, and DC + DT represents their combination.
CDSSM-ft refers to fine-tuning using labeled data after training CDSSM on scored, unlabeled samples.
As mentioned in Section~\ref{sec.introduction}, training by clicked data means taking clicked $<$\textit{query}, \textit{ad}$>$ pair as positive, and randomly synthetic query and listing pairs as negative.

The first row in Table~\ref{table.comparison} presents the results of our click and labeled baseline, obtained by training CDSSM purely on user clicks and relevance labels.
% Compared with the click baseline, the labeled baseline achieves an improvement of 13.17\% and 10.4\% in terms of \textit{ROC AUC} and \textit{PR AUC} respectively, which validates our analysis that training by user clicks introduces severe ambiguities.
The next row gives results of our annotation models, where DC and DT achieves comparable performance. Their combination brings further improvements. 
% As previously mentioned, both annotation models cannot be deployed as fast matching models since they are either inseparable or depends on inaccessible features.
Training by the proposed framework with DC as annotation model outperforms the click baseline by 9.8\% for \textit{ROC AUC}, and 9.32\% by \textit{PR AUC}. Very close result is observed for DT.
Learning by the proposed framework achieves comparable performance to the labeled baseline when combining DT and DC.
Finally, when further fine-tuned on labeled data, the \textit{ROC AUC} using annotation model DC and DT outperforms the labeled baseline by 2.6\% and 3.82\% respectively.
Combining DC and DT as annotation model outperforms the labeled baseline by a large margin of 4.5\% for \textit{ROC AUC}, and 3.2\% for \textit{PR AUC}.
%More importantly, combining DC and DT as teacher model achieves an overall \textit{ROC AUC} as high as $0.8497$, outperforming the labeled baseline by a large margin of 4.5\%.
This demonstrates evidently the effectiveness of the proposed framework.

\subsection{Annotation Models and Mapping Functions}
\label{sec.exp.influence}

Several configurations of annotation models and mapping functions are compared, with results summarized in Table~\ref{table.analyze}.
Specifically, the annotation model is implemented as several variants of Deep Crossing, including the original version, Deep Crossing with Batch Normalization~(BN), Deep Crossing with BN and Attention~(abbreviated as A), Deep Crossing with BN and MTL, and Deep Crossing with BN, attention and MTL.
For each annotation model, three mapping functions are implemented, and \textit{AUC} before and after fine-tuning are reported, with $\theta = 1$.

As shown in Table~\ref{table.analyze}, the performance of CDSSM are highly correlated with that of annotation models.
%: the better the teacher model is, the better the student model is.
% This is particularly preferable in practice, since improving annotation models is much easier than directly making improvements on CDSSM, and will not introduce any additional computations in the online serving phase.
%For instance, we can add human designed rules, use model ensembles, and introduce more complex model architectures or feature engineering, etc.
In addition, Table~\ref{table.analyze} also demonstrates that training annotation models by jointly learning multiple tasks could bring about considerable gains in both annotation models and CDSSM, despite its simplicity. 
% The method used for constructing and learning related tasks here could potentially benefit other tasks involving enumerated labels.
Table~\ref{table.analyze} also reveals the importance of mapping functions, where the third configuration outperforms others in most cases, across different annotation models. This may due to the reason similar to prior analysis in~\cite{hinton2015distilling}, that the soft targets could offer more information than hard targets.

\subsection{Label-aware Weights}
\label{sec.exp.weight}

\begin{table}[!t]
\caption{Comparison of different fine-tuning methods.}
    \begin{center}
    \begin{tabular}{l|l|c|c}
        \thickhline
                                        %&           &\multicolumn{2}{|c}{\textit{AUC}}    \\
        %\cline{3-4}
                                        &           &\textit{ROC AUC}   &\textit{PR AUC}\\
        \hline
        \multirow{2}{*}{baseline}       &hard       &0.8271             &0.8833\\
                                        &soft       &0.8372             &0.8922\\
        \hline
        \multirow{4}{*}{$\theta$}       &0.0        &0.8390             &0.8935\\
                                        &0.2        &0.8391             &0.8935\\
                                        &0.5        &\bf{0.8394}        &0.8939\\
                                        &0.8        &0.8378             &0.8926\\
        \thickhline
    \end{tabular}
    \end{center}
    \label{table.theta}
\end{table}

To show the influence of label-aware weights, we conduct experiments on DC-BN-A-MTL by setting $\theta$ to be~\{0, 0.2, 0.5, 0.8, 1.0\}, with results summarized in Table~\ref{table.theta}, where the hard baseline refers to directly learning binary label $\tilde{y}$, and the soft baseline refers to learning target $y$.
%and the merged baseline refers to learning the arithmetic mean of $\tilde{y}$ and $y$, all
% both by minimizing the corresponding weighted MSE loss.
% Note that the soft baseline equivalents to setting $\theta = 1$.

According to Table~\ref{table.theta}, fine-tuning by the hard baseline yields suboptimal performance, which coincides with prior analysis in~\cite{hinton2015distilling}.
In addition, the best \textit{AUC} is achieved at $\theta = 0.5$, which equivalents to setting the weights for unacceptable loss twice as much as that of the acceptable ones.
Compared with the case $\theta = 0.5$, both $\theta = 0$ and $\theta = 1$ give inferior performance.
More specifically, $\theta = 1$ exhibits a larger gap with $\theta = 0.5$, due to its inability to leverage human provided labels.
In contrast, $\theta = 0$ simply spares all acceptable loss, and hence has limited influence on the overall performance.
% Based upon the above analysis, the proposed label-aware weights are relatively insensitive to $\theta$ when it approaches zero, and becomes more sensitive as $\theta$ grows.

\subsection{Improve Data Efficiency}
\label{sec.exp.robust}

\begin{figure}[!t]
\begin{center}
   \includegraphics[width=0.5\textwidth]{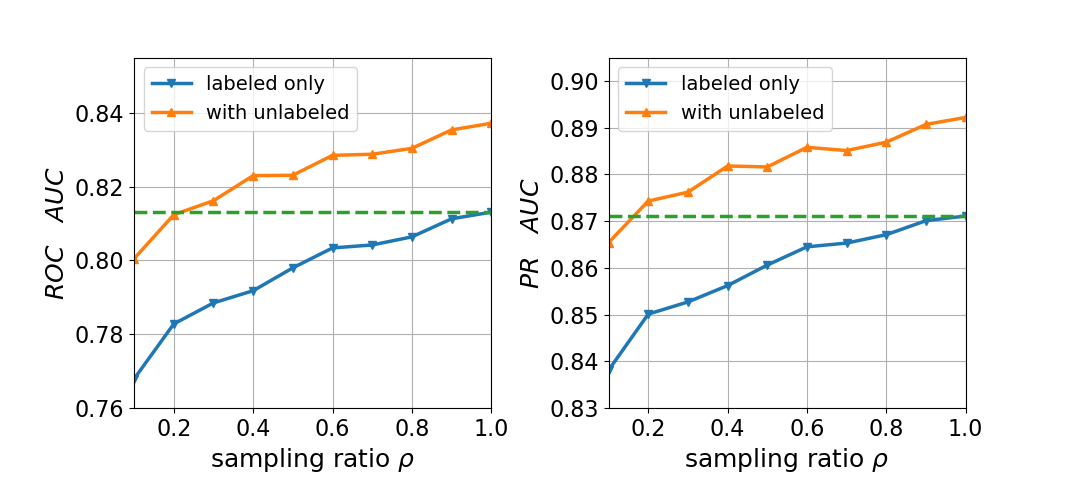}
\end{center}
   \caption{Performance of sampling labeled data by different ratios. Blue curves represent training on the sampled data, yellow curves represent training by the proposed method.}
   \label{fig.robust}
\end{figure}

This section investigates the potential of complementing labeled data using unlabeled samples.
To figure out this, we randomly sample 9 subsets from the labeled data using a sampling ratio $\rho$ ranging from 0.1 to 0.9, and repeat the steps in Figure~\ref{fig.overview} using the sampled labeled data only.
The results are shown in Figure~\ref{fig.robust}, where the blue curves show the \textit{AUC} obtained by training CDSSM directly on the sampled labeled data, while the yellow curves denote performance achieved by following the proposed framework.
% , i.e., training Deep Crossing on the sampled labeled data, scoring unlabeled data using Deep Crossing, training CDSSM on the scored unlabeled data, and fine-tuning CDSSM using the sampled labeled data.

According to Figure~\ref{fig.robust}, all \textit{AUC} curves grow monotonically as $\rho$ gets larger, but both \textit{ROC AUC} and \textit{PR AUC} grows slower and slower, indicating decreasing data efficiency.
% More precisely, the \textit{ROC AUC} becomes 0.7679 with 10\% labeled data, however the remaining 90\% data only improves it to 0.8131.
The green dashed lines in Figure~\ref{fig.robust} highlight the upper bound of the blue curves, which is easily exceeded when training using the proposed framework: the same \textit{AUC} is achieved by the orange curve with only 20\% labeled data.
This validates that training by the proposed framework could greatly improve data efficiency. With 80\% knowledge could be obtained from unlabeled data, our dependency on the expensive human labels could be greatly reduced.

% -----------------------------------------------------------------------------------------------------------------------------------------------------------------------------------------

\section{Conclusions}
\label{sec.conclusion}

This paper proposes a novel framework for learning the relevance model in sponsored search, allowing us to learn a simple model from stronger but undeployable annotators, and offers a principled way to leverage unlabeled data by completely dispenses clicks.
% Specifically, the proposed framework consists of four major steps.
% In the first step, we construct a bunch of auxiliary tasks based upon the enumerated relevance labels, and train several annotation models by jointly learning multiple tasks.
% The learnt annotation models are then used to assign scores to both the unlabeled and labeled samples.
% Then, a simple and deployable model is trained using scored unlabeled data, and fine-tuned using the scored labeled data, by leveraging both weak annotations and true human labels via minimizing the proposed label-aware weighted loss.
Our main experimental results include 1) training by the proposed framework could substantially improve the performance of relevance models, and 2) the proposed framework is able to significantly increase data efficiency, so that training by 20\% labeled data could achieve comparable results with the baseline trained by all labeled samples. 
% In conclusion, our experiments demonstrate that the proposed framework could greatly benefit the learning of relevance model in sponsored search.

%
% The next two lines define the bibliography style to be used, and the bibliography file.
\bibliographystyle{ACM-Reference-Format}
\bibliography{bibfile}


\begin{thebibliography}{31}

%%% the Web address.

\ifx \showCODEN    \undefined \def \showCODEN     #1{\unskip}     \fi
\ifx \showDOI      \undefined \def \showDOI       #1{#1}\fi
\ifx \showISBNx    \undefined \def \showISBNx     #1{\unskip}     \fi
\ifx \showISBNxiii \undefined \def \showISBNxiii  #1{\unskip}     \fi
\ifx \showISSN     \undefined \def \showISSN      #1{\unskip}     \fi
\ifx \showLCCN     \undefined \def \showLCCN      #1{\unskip}     \fi
\ifx \shownote     \undefined \def \shownote      #1{#1}          \fi
\ifx \showarticletitle \undefined \def \showarticletitle #1{#1}   \fi
\ifx \showURL      \undefined \def \showURL       {\relax}        \fi
% invisible to TeX
\providecommand\bibfield[2]{#2}
\providecommand\bibinfo[2]{#2}
\providecommand\natexlab[1]{#1}
\providecommand\showeprint[2][]{arXiv:#2}

\bibitem[\protect\citeauthoryear{Ba and Caruana}{Ba and Caruana}{2014}]%
        {ba2014do}
\bibfield{author}{\bibinfo{person}{Jimmy Ba} {and} \bibinfo{person}{Rich
  Caruana}.} \bibinfo{year}{2014}\natexlab{}.
\newblock \showarticletitle{Do Deep Nets Really Need to be Deep?}
\newblock In \bibinfo{booktitle}{\emph{Advances in Neural Information
  Processing Systems}}. \bibinfo{pages}{2654--2662}.
\newblock


\bibitem[\protect\citeauthoryear{Bahdanau, Cho, and Bengio}{Bahdanau
  et~al\mbox{.}}{2014}]%
        {bahdanau2014neural}
\bibfield{author}{\bibinfo{person}{Dzmitry Bahdanau},
  \bibinfo{person}{Kyunghyun Cho}, {and} \bibinfo{person}{Yoshua Bengio}.}
  \bibinfo{year}{2014}\natexlab{}.
\newblock \bibinfo{title}{Neural Machine Translation by Jointly Learning to
  Align and Translate}.  (\bibinfo{year}{2014}).
\newblock
\showeprint[arxiv]{1409.0473}
\urldef\tempurl%
\url{http://arxiv.org/abs/1409.0473}
\showURL{%
\tempurl}


\bibitem[\protect\citeauthoryear{Blei, Ng, and Jordan}{Blei
  et~al\mbox{.}}{2003}]%
        {blei2003latent}
\bibfield{author}{\bibinfo{person}{David~M. Blei}, \bibinfo{person}{Andrew~Y.
  Ng}, {and} \bibinfo{person}{Michael~I. Jordan}.}
  \bibinfo{year}{2003}\natexlab{}.
\newblock \showarticletitle{Latent Dirichlet Allocation}.
\newblock \bibinfo{journal}{\emph{Journal of Machine Learning Research}}
  \bibinfo{volume}{3} (\bibinfo{year}{2003}), \bibinfo{pages}{993--1022}.
\newblock


\bibitem[\protect\citeauthoryear{Bucilu\v{a}, Caruana, and
  Niculescu-Mizil}{Bucilu\v{a} et~al\mbox{.}}{2006}]%
        {bucilua2006model}
\bibfield{author}{\bibinfo{person}{Cristian Bucilu\v{a}}, \bibinfo{person}{Rich
  Caruana}, {and} \bibinfo{person}{Alexandru Niculescu-Mizil}.}
  \bibinfo{year}{2006}\natexlab{}.
\newblock \showarticletitle{Model Compression}. In
  \bibinfo{booktitle}{\emph{International Conference on Knowledge Discovery and
  Data Mining}}. \bibinfo{publisher}{ACM}, \bibinfo{pages}{535--541}.
\newblock


\bibitem[\protect\citeauthoryear{Deerwester, Dumais, Furnas, Landauer, and
  Harshman}{Deerwester et~al\mbox{.}}{1990}]%
        {deerwester1990lsa}
\bibfield{author}{\bibinfo{person}{Scott Deerwester}, \bibinfo{person}{Susan~T.
  Dumais}, \bibinfo{person}{George~W. Furnas}, \bibinfo{person}{Thomas~K.
  Landauer}, {and} \bibinfo{person}{Richard Harshman}.}
  \bibinfo{year}{1990}\natexlab{}.
\newblock \showarticletitle{Indexing by Latent Semantic Analysis}.
\newblock \bibinfo{journal}{\emph{Journal of the American Society for
  Information Science}}  \bibinfo{volume}{41} (\bibinfo{year}{1990}),
  \bibinfo{pages}{391--407}.
\newblock


\bibitem[\protect\citeauthoryear{Gao, Pantel, Gamon, He, Deng, and Shen}{Gao
  et~al\mbox{.}}{2014}]%
        {gao2014modeling}
\bibfield{author}{\bibinfo{person}{Jianfeng Gao}, \bibinfo{person}{Patrick
  Pantel}, \bibinfo{person}{Michael Gamon}, \bibinfo{person}{Xiaodong He},
  \bibinfo{person}{Li Deng}, {and} \bibinfo{person}{Yelong Shen}.}
  \bibinfo{year}{2014}\natexlab{}.
\newblock \showarticletitle{Modeling Interestingness with Deep Neural
  Networks}. In \bibinfo{booktitle}{\emph{Conference on Empirical Methods in
  Natural Language Processing}}. \bibinfo{pages}{2--13}.
\newblock


\bibitem[\protect\citeauthoryear{Gao, Toutanova, and Yih}{Gao
  et~al\mbox{.}}{2011}]%
        {gao2011clickthrough}
\bibfield{author}{\bibinfo{person}{Jianfeng Gao}, \bibinfo{person}{Kristina
  Toutanova}, {and} \bibinfo{person}{Wen-tau Yih}.}
  \bibinfo{year}{2011}\natexlab{}.
\newblock \showarticletitle{Clickthrough-based Latent Semantic Models for Web
  Search}. In \bibinfo{booktitle}{\emph{International SIGIR Conference on
  Research and Development in Information Retrieval}}
  \emph{(\bibinfo{series}{SIGIR '11})}. \bibinfo{publisher}{ACM},
  \bibinfo{pages}{675--684}.
\newblock


\bibitem[\protect\citeauthoryear{Guo, Fan, Ai, and Croft}{Guo
  et~al\mbox{.}}{2016}]%
        {guo2016deep}
\bibfield{author}{\bibinfo{person}{Jiafeng Guo}, \bibinfo{person}{Yixing Fan},
  \bibinfo{person}{Qingyao Ai}, {and} \bibinfo{person}{W.~Bruce Croft}.}
  \bibinfo{year}{2016}\natexlab{}.
\newblock \showarticletitle{A Deep Relevance Matching Model for Ad-hoc
  Retrieval}. In \bibinfo{booktitle}{\emph{International Conference on
  Information and Knowledge Management}}. \bibinfo{publisher}{ACM},
  \bibinfo{pages}{55--64}.
\newblock


\bibitem[\protect\citeauthoryear{{Hinton}, {Vinyals}, and {Dean}}{{Hinton}
  et~al\mbox{.}}{2015}]%
        {hinton2015distilling}
\bibfield{author}{\bibinfo{person}{G. {Hinton}}, \bibinfo{person}{O.
  {Vinyals}}, {and} \bibinfo{person}{J. {Dean}}.}
  \bibinfo{year}{2015}\natexlab{}.
\newblock \bibinfo{title}{Distilling the Knowledge in a Neural Network}.
  (\bibinfo{date}{March} \bibinfo{year}{2015}).
\newblock
\showeprint[arxiv]{stat.ML/1503.02531}


\bibitem[\protect\citeauthoryear{Hofmann}{Hofmann}{1999}]%
        {hofmann1999plsa}
\bibfield{author}{\bibinfo{person}{Thomas Hofmann}.}
  \bibinfo{year}{1999}\natexlab{}.
\newblock \showarticletitle{Probabilistic Latent Semantic Analysis}. In
  \bibinfo{booktitle}{\emph{Conference on Uncertainty in Artificial
  Intelligence}}. \bibinfo{publisher}{Morgan Kaufmann Publishers Inc.},
  \bibinfo{pages}{289--296}.
\newblock


\bibitem[\protect\citeauthoryear{Hu, Lu, Li, and Chen}{Hu
  et~al\mbox{.}}{2014}]%
        {hu2014convolutional}
\bibfield{author}{\bibinfo{person}{Baotian Hu}, \bibinfo{person}{Zhengdong Lu},
  \bibinfo{person}{Hang Li}, {and} \bibinfo{person}{Qingcai Chen}.}
  \bibinfo{year}{2014}\natexlab{}.
\newblock \showarticletitle{Convolutional Neural Network Architectures for
  Matching Natural Language Sentences}.
\newblock In \bibinfo{booktitle}{\emph{Advances in Neural Information
  Processing Systems}}. \bibinfo{publisher}{Curran Associates, Inc.},
  \bibinfo{pages}{2042--2050}.
\newblock


\bibitem[\protect\citeauthoryear{Huang, He, Gao, Deng, Acero, and Heck}{Huang
  et~al\mbox{.}}{2013}]%
        {huang2013learning}
\bibfield{author}{\bibinfo{person}{Po-Sen Huang}, \bibinfo{person}{Xiaodong
  He}, \bibinfo{person}{Jianfeng Gao}, \bibinfo{person}{Li Deng},
  \bibinfo{person}{Alex Acero}, {and} \bibinfo{person}{Larry Heck}.}
  \bibinfo{year}{2013}\natexlab{}.
\newblock \showarticletitle{Learning Deep Structured Semantic Models for Web
  Search Using Clickthrough Data}. In \bibinfo{booktitle}{\emph{International
  Conference on Information and Knowledge Management}}. ACM,
  \bibinfo{pages}{2333--2338}.
\newblock


\bibitem[\protect\citeauthoryear{Ioffe and Szegedy}{Ioffe and Szegedy}{2015}]%
        {sergey2015batch}
\bibfield{author}{\bibinfo{person}{Sergey Ioffe} {and}
  \bibinfo{person}{Christian Szegedy}.} \bibinfo{year}{2015}\natexlab{}.
\newblock \bibinfo{title}{Batch Normalization: Accelerating Deep Network
  Training by Reducing Internal Covariate Shift}.  (\bibinfo{year}{2015}).
\newblock
\showeprint[arxiv]{1502.03167}
\urldef\tempurl%
\url{http://arxiv.org/abs/1502.03167}
\showURL{%
\tempurl}


\bibitem[\protect\citeauthoryear{Ling, Deng, Gu, Zhou, Li, and Sun}{Ling
  et~al\mbox{.}}{2017}]%
        {ling2017model}
\bibfield{author}{\bibinfo{person}{Xiaoliang Ling}, \bibinfo{person}{Weiwei
  Deng}, \bibinfo{person}{Chen Gu}, \bibinfo{person}{Hucheng Zhou},
  \bibinfo{person}{Cui Li}, {and} \bibinfo{person}{Feng Sun}.}
  \bibinfo{year}{2017}\natexlab{}.
\newblock \showarticletitle{Model Ensemble for Click Prediction in Bing Search
  Ads}. In \bibinfo{booktitle}{\emph{International Conference on World Wide Web
  Companion}}. \bibinfo{pages}{689--698}.
\newblock


\bibitem[\protect\citeauthoryear{Lu and Li}{Lu and Li}{2013}]%
        {lu2013advances}
\bibfield{author}{\bibinfo{person}{Zhengdong Lu} {and} \bibinfo{person}{Hang
  Li}.} \bibinfo{year}{2013}\natexlab{}.
\newblock \showarticletitle{A Deep Architecture for Matching Short Texts}.
\newblock In \bibinfo{booktitle}{\emph{Advances in Neural Information
  Processing Systems}}, \bibfield{editor}{\bibinfo{person}{C.~J.~C. Burges},
  \bibinfo{person}{L.~Bottou}, \bibinfo{person}{M.~Welling},
  \bibinfo{person}{Z.~Ghahramani}, {and} \bibinfo{person}{K.~Q. Weinberger}}
  (Eds.). \bibinfo{publisher}{Curran Associates, Inc.},
  \bibinfo{pages}{1367--1375}.
\newblock
\urldef\tempurl%
\url{http://papers.nips.cc/paper/5019-a-deep-architecture-for-matching-short-texts.pdf}
\showURL{%
\tempurl}


\bibitem[\protect\citeauthoryear{Mitra and Craswell}{Mitra and
  Craswell}{2017}]%
        {mitra2017neural}
\bibfield{author}{\bibinfo{person}{Bhaskar Mitra} {and} \bibinfo{person}{Nick
  Craswell}.} \bibinfo{year}{2017}\natexlab{}.
\newblock \bibinfo{title}{Neural Models for Information Retrieval}.
  (\bibinfo{year}{2017}).
\newblock
\showeprint[arxiv]{1705.01509}
\urldef\tempurl%
\url{http://arxiv.org/abs/1705.01509}
\showURL{%
\tempurl}


\bibitem[\protect\citeauthoryear{Mitra, Diaz, and Craswell}{Mitra
  et~al\mbox{.}}{2017}]%
        {mitra2017learning}
\bibfield{author}{\bibinfo{person}{Bhaskar Mitra}, \bibinfo{person}{Fernando
  Diaz}, {and} \bibinfo{person}{Nick Craswell}.}
  \bibinfo{year}{2017}\natexlab{}.
\newblock \showarticletitle{Learning to Match Using Local and Distributed
  Representations of Text for Web Search}. In
  \bibinfo{booktitle}{\emph{International Conference on World Wide Web}}.
  \bibinfo{pages}{1291--1299}.
\newblock


\bibitem[\protect\citeauthoryear{Parikh, T{\"{a}}ckstr{\"{o}}m, Das, and
  Uszkoreit}{Parikh et~al\mbox{.}}{2016}]%
        {parikh2016decomposable}
\bibfield{author}{\bibinfo{person}{Ankur~P. Parikh}, \bibinfo{person}{Oscar
  T{\"{a}}ckstr{\"{o}}m}, \bibinfo{person}{Dipanjan Das}, {and}
  \bibinfo{person}{Jakob Uszkoreit}.} \bibinfo{year}{2016}\natexlab{}.
\newblock \bibinfo{title}{A Decomposable Attention Model for Natural Language
  Inference}.  (\bibinfo{year}{2016}).
\newblock
\showeprint[arxiv]{1606.01933}
\urldef\tempurl%
\url{http://arxiv.org/abs/1606.01933}
\showURL{%
\tempurl}


\bibitem[\protect\citeauthoryear{Romero, Ballas, Kahou, Chassang, Gatta, and
  Bengio}{Romero et~al\mbox{.}}{2014}]%
        {romero2014}
\bibfield{author}{\bibinfo{person}{Adriana Romero}, \bibinfo{person}{Nicolas
  Ballas}, \bibinfo{person}{Samira~Ebrahimi Kahou}, \bibinfo{person}{Antoine
  Chassang}, \bibinfo{person}{Carlo Gatta}, {and} \bibinfo{person}{Yoshua
  Bengio}.} \bibinfo{year}{2014}\natexlab{}.
\newblock \bibinfo{title}{FitNets: Hints for Thin Deep Nets}.
  (\bibinfo{year}{2014}).
\newblock
\showeprint[arxiv]{1412.6550}
\urldef\tempurl%
\url{http://arxiv.org/abs/1412.6550}
\showURL{%
\tempurl}


\bibitem[\protect\citeauthoryear{Salakhutdinov and Hinton}{Salakhutdinov and
  Hinton}{2009}]%
        {salakhutdinov2009semantic}
\bibfield{author}{\bibinfo{person}{Ruslan Salakhutdinov} {and}
  \bibinfo{person}{Geoffrey Hinton}.} \bibinfo{year}{2009}\natexlab{}.
\newblock \showarticletitle{Semantic Hashing}.
\newblock \bibinfo{journal}{\emph{International Journal of Approximate
  Reasoning}} \bibinfo{volume}{50}, \bibinfo{number}{7} (\bibinfo{year}{2009}),
  \bibinfo{pages}{969 -- 978}.
\newblock


\bibitem[\protect\citeauthoryear{Sau and Balasubramanian}{Sau and
  Balasubramanian}{2016}]%
        {bharat2016deep}
\bibfield{author}{\bibinfo{person}{Bharat~Bhusan Sau} {and}
  \bibinfo{person}{Vineeth~N. Balasubramanian}.}
  \bibinfo{year}{2016}\natexlab{}.
\newblock \bibinfo{title}{Deep Model Compression: Distilling Knowledge from
  Noisy Teachers}.  (\bibinfo{year}{2016}).
\newblock
\showeprint[arxiv]{1610.09650}
\urldef\tempurl%
\url{http://arxiv.org/abs/1610.09650}
\showURL{%
\tempurl}


\bibitem[\protect\citeauthoryear{Severyn and Moschitti}{Severyn and
  Moschitti}{2015}]%
        {severyn2015learning}
\bibfield{author}{\bibinfo{person}{Aliaksei Severyn} {and}
  \bibinfo{person}{Alessandro Moschitti}.} \bibinfo{year}{2015}\natexlab{}.
\newblock \showarticletitle{Learning to Rank Short Text Pairs with
  Convolutional Deep Neural Networks}. In
  \bibinfo{booktitle}{\emph{International Conference on Research and
  Development in Information Retrieval}}. \bibinfo{publisher}{ACM},
  \bibinfo{pages}{373--382}.
\newblock


\bibitem[\protect\citeauthoryear{Shan, Hoens, Jiao, Wang, Yu, and Mao}{Shan
  et~al\mbox{.}}{2016}]%
        {shan2016deep}
\bibfield{author}{\bibinfo{person}{Ying Shan}, \bibinfo{person}{T.~Ryan Hoens},
  \bibinfo{person}{Jian Jiao}, \bibinfo{person}{Haijing Wang},
  \bibinfo{person}{Dong Yu}, {and} \bibinfo{person}{JC Mao}.}
  \bibinfo{year}{2016}\natexlab{}.
\newblock \showarticletitle{Deep Crossing: Web-Scale Modeling Without Manually
  Crafted Combinatorial Features}. In \bibinfo{booktitle}{\emph{International
  Conference on Knowledge Discovery and Data Mining}}.
  \bibinfo{publisher}{ACM}, \bibinfo{pages}{255--262}.
\newblock


\bibitem[\protect\citeauthoryear{Shen, He, Gao, Deng, and Mesnil}{Shen
  et~al\mbox{.}}{2014a}]%
        {shen2014latent}
\bibfield{author}{\bibinfo{person}{Yelong Shen}, \bibinfo{person}{Xiaodong He},
  \bibinfo{person}{Jianfeng Gao}, \bibinfo{person}{Li Deng}, {and}
  \bibinfo{person}{Gr{\'e}goire Mesnil}.} \bibinfo{year}{2014}\natexlab{a}.
\newblock \showarticletitle{A Latent Semantic Model with Convolutional-Pooling
  Structure for Information Retrieval}. In
  \bibinfo{booktitle}{\emph{International Conference on Information and
  Knowledge Management}}. \bibinfo{publisher}{ACM}, \bibinfo{pages}{101--110}.
\newblock


\bibitem[\protect\citeauthoryear{Shen, He, Gao, Deng, and Mesnil}{Shen
  et~al\mbox{.}}{2014b}]%
        {shen2014learning}
\bibfield{author}{\bibinfo{person}{Yelong Shen}, \bibinfo{person}{Xiaodong He},
  \bibinfo{person}{Jianfeng Gao}, \bibinfo{person}{Li Deng}, {and}
  \bibinfo{person}{Gr{\'e}goire Mesnil}.} \bibinfo{year}{2014}\natexlab{b}.
\newblock \showarticletitle{Learning Semantic Representations using
  Convolutional Neural Networks for Web Search}. In
  \bibinfo{booktitle}{\emph{International Conference on World Wide Web}}.
  \bibinfo{publisher}{ACM}, \bibinfo{pages}{373--374}.
\newblock


\bibitem[\protect\citeauthoryear{Tai, Socher, and Manning}{Tai
  et~al\mbox{.}}{2015}]%
        {tai2015improved}
\bibfield{author}{\bibinfo{person}{Kai~Sheng Tai}, \bibinfo{person}{Richard
  Socher}, {and} \bibinfo{person}{Christopher~D. Manning}.}
  \bibinfo{year}{2015}\natexlab{}.
\newblock \bibinfo{title}{Improved Semantic Representations From
  Tree-Structured Long Short-Term Memory Networks}.  (\bibinfo{year}{2015}).
\newblock
\showeprint[arxiv]{1503.00075}
\urldef\tempurl%
\url{http://arxiv.org/abs/1503.00075}
\showURL{%
\tempurl}


\bibitem[\protect\citeauthoryear{Wan, Lan, Guo, Xu, Pang, and Cheng}{Wan
  et~al\mbox{.}}{2016}]%
        {wan2016deep}
\bibfield{author}{\bibinfo{person}{Shengxian Wan}, \bibinfo{person}{Yanyan
  Lan}, \bibinfo{person}{Jiafeng Guo}, \bibinfo{person}{Jun Xu},
  \bibinfo{person}{Liang Pang}, {and} \bibinfo{person}{Xueqi Cheng}.}
  \bibinfo{year}{2016}\natexlab{}.
\newblock \showarticletitle{A Deep Architecture for Semantic Matching with
  Multiple Positional Sentence Representations}. In
  \bibinfo{booktitle}{\emph{AAAI Conference on Artificial Intelligence}}.
  \bibinfo{pages}{2835--2841}.
\newblock


\bibitem[\protect\citeauthoryear{Wang and Li}{Wang and Li}{2012}]%
        {wang2012query}
\bibfield{author}{\bibinfo{person}{Jingdong Wang} {and}
  \bibinfo{person}{Shipeng Li}.} \bibinfo{year}{2012}\natexlab{}.
\newblock \showarticletitle{Query-driven Iterated Neighborhood Graph Search for
  Large Scale Indexing}. In \bibinfo{booktitle}{\emph{International Conference
  on Multimedia}}. ACM, \bibinfo{pages}{179--188}.
\newblock


\bibitem[\protect\citeauthoryear{Wu, Wu, Li, and Zhou}{Wu
  et~al\mbox{.}}{2018}]%
        {wu018learning}
\bibfield{author}{\bibinfo{person}{Yu Wu}, \bibinfo{person}{Wei Wu},
  \bibinfo{person}{Zhoujun Li}, {and} \bibinfo{person}{Ming Zhou}.}
  \bibinfo{year}{2018}\natexlab{}.
\newblock \bibinfo{title}{Learning Matching Models with Weak Supervision for
  Response Selection in Retrieval-based Chatbots}.  (\bibinfo{year}{2018}).
\newblock
\showeprint[arxiv]{1805.02333}
\urldef\tempurl%
\url{http://arxiv.org/abs/1805.02333}
\showURL{%
\tempurl}


\bibitem[\protect\citeauthoryear{Yang, Ai, Guo, and Croft}{Yang
  et~al\mbox{.}}{2016}]%
        {yang2016anmm}
\bibfield{author}{\bibinfo{person}{Liu Yang}, \bibinfo{person}{Qingyao Ai},
  \bibinfo{person}{Jiafeng Guo}, {and} \bibinfo{person}{W.~Bruce Croft}.}
  \bibinfo{year}{2016}\natexlab{}.
\newblock \showarticletitle{aNMM: Ranking Short Answer Texts with
  Attention-Based Neural Matching Model}. In
  \bibinfo{booktitle}{\emph{International Conference on Information and
  Knowledge Management}}. \bibinfo{publisher}{ACM}, \bibinfo{pages}{287--296}.
\newblock


\bibitem[\protect\citeauthoryear{Yin, Schütze, Xiang, and Zhou}{Yin
  et~al\mbox{.}}{2016}]%
        {yin2016abcnn}
\bibfield{author}{\bibinfo{person}{Wenpeng Yin}, \bibinfo{person}{Hinrich
  Schütze}, \bibinfo{person}{Bing Xiang}, {and} \bibinfo{person}{Bowen Zhou}.}
  \bibinfo{year}{2016}\natexlab{}.
\newblock \showarticletitle{ABCNN: Attention-Based Convolutional Neural Network
  for Modeling Sentence Pairs}.
\newblock \bibinfo{journal}{\emph{Transactions of the Association for
  Computational Linguistics}}  \bibinfo{volume}{4} (\bibinfo{year}{2016}),
  \bibinfo{pages}{259 -- 272}.
\newblock


\end{thebibliography}

\end{document}